\documentclass[a4paper,11pt]{article}
\usepackage{amsmath}
\usepackage{amsfonts}
\usepackage{amssymb}
\usepackage[dvips]{graphicx}
\providecommand{\U}[1]{\protect\rule{.1in}{.1in}}
\providecommand{\U}[1]{\protect\rule{.1in}{.1in}}
\newtheorem{theorem}{Theorem}

\newtheorem{definition}[theorem]{Definition}

\newtheorem{idea memo}[theorem]{Idea Memo}

\newtheorem{proposition}[theorem]{Proposition}

\newcommand{\mapright}[1]{\smash{\mathop{\hbox to 1cm{\rightarrowfill}}\limits^{#1}}}
\newcommand{\mapleft}[1]{\smash{\mathop{\hbox to 1cm{\leftarrowfill}}\limits^{#1}}}

\begin{document}

\title{Operational and harmonic-analytic aspects of quasi-probability distributions}
\author{Ryo HARADA\footnote{rharada@kurims.kyoto-u.ac.jp}\\Research Institute for Mathematical Sciences,\\Kyoto University, Kyoto 606-8502, Japan}
\date{\today}
\maketitle

\begin{abstract}
Husimi distributions \cite{Husimi} and Wigner distributions \cite{Wigner} are well-known quasi-probability distributions which appear in several contexts. In this paper, we show some remarkable aspects of these distribution functions related to geometric structures of generalized coherent state systems \cite{Per,Per2} and operational quantum physics \cite{Busch}, and a scheme of formulating generalized version of quasi-probability distributions. Our scheme gives concrete formulae of quasi-probability distributions in more direct way from the theory of coherent state systems and clarify their operational meanings, especially of Husimi distributions and mutual relation between Husimi distributions and other classes of quasi-probability distributions.
\end{abstract}

\section{Introduction}
For the purpose of obtaining information of microscopic world, it is necessary to examine some channels which connect microscopic systems with macroscopic ingredients. On the basis of such a viewpoint, many researches categorized as {\em quantum measurement theory} have been done so far and some are in progress. In typical cases, quantum measurement processes can be formulated as construction of some (probabilistic) measure over macroscopic order parameters as integration of measurement results. More physically, it is natural to suppose some macroscopic phase space whose coordinate variables are expectation values of observables in the system under consideration.

In this paper, we focus on quantum systems with Lie-group symmetries. In such cases, the theory of generalized coherent state systems (GCS systems, in short; see \cite{Kla,Per}, for example) is well known as a useful tool owing to the existence of coherent phase spaces with meaningful physical structures. On the basis of this framework we can construct some systematic formulation of quasi-probability distributions (QPD's, in short) including generalized Wigner and Husimi distributions. We can find some technical hints for our formulation of QPD's in the discussions given by V\'{a}rilly and Gracia-Band\'{i}a for $SU(2)$ group \cite{V-G} or Brif and Mann for arbitrary homogeneous spaces with Lie-group symmetries \cite{Brif-Mann}.

A remarkable application of QPD's is comprehension of quantum systems on (semi)classical phase spaces: Let us consider a dynamical equation of some quantum systems
\begin{equation}\label{DynEq}
\frac{\partial\rho}{\partial t}=\mathcal{P}(\rho,\hat{A_{1}},\hat{A_{2}},\cdots)\mbox{\qquad}(t\in\mathbb{R}),
\end{equation}
where $\mathcal{P}$ is some polynomial and $A_{1}$, $A_{2}$, and a density operator $\rho$ belong to a $*$-algebra $\mathcal{A}$ (for example, quantum Master equations such as $\frac{\partial\rho}{\partial t}=\sum_{m,n}T_{m,n}[\hat{A}_{m}\rho,\hat{A}_{n}]+h.c.$ ($T_{m,n}\in\mathbb{C}$) in contexts of relaxation processes \cite{Ban}). Our formulation shows us the positivity for resolving these problems into differential equations of functions on some phase space; GCS systems contain basic ingredients for this goal.

The essence of this paper is summarized in the following two results which are mutually related:
\begin{itemize}
\item We have found an operational interpretation of (generalized) Husimi distributions as a minimizer of uncertainty.
\item We have formulated a scheme to define QPD's completely in terms of GCS systems.
\end{itemize}
Namely, GCS systems are understood to have enough information for construction of QPD's. These results show the importance of the idea of coherent states, especially in operational quantum physics and quantum probability theory, which implies possibility of broad applications, for example, description of dynamical equations of quantum systems via QPD's as mentioned above. Some central notions of GCS systems are shown in \S\ref{GeneralTheory}.

\section{Complex geometric structures in coherent state systems}

In this section we review the essence of the theory of GCS systems \cite{Kla,Per} and prepare the required properties and notations for our discussion, especially the idea of {\em coherent phase spaces} with symplectic (K\"{a}hler) geometric structures and representation of states on them.

\subsection{Generalized coherent state systems}\label{GeneralTheory}

Let $G$ be a real Lie group and $U$ be one of its unitary irreducible representations on a Hilbert space $\mathcal{H}$. Fix an arbitrary vector ${\psi}_{0}\in\mathcal{H}$. Then we can consider the $G$-orbit starting from $|\psi_{0}\rangle$ in $\mathbb{P}(\mathcal{H})$ (: the projective space associated to $\mathcal{H}$) as follows:
\[
\{|{\psi}_{g}\rangle :=U(g)|{\psi}_{0}\rangle ;g\in G \}.
\]
In order to clarify characteristic features of this orbit, we consider an important subgroup $H<G$, called an isotropy subgroup with respect to ${\psi}_{0}\in\mathcal{H}$. The isotropy subgroup $H$ is defined as
\[
H:=\{h\in G ; \exists\alpha :H\to \mathbb{R},U(h)|{\psi}_{0}\rangle=\exp (i\alpha (h))|{\psi}_{0}\rangle \}.
\]
Then we can find a 1-to-1 correspondence between the $G$-orbit $\{ |{\psi}_{g}\rangle ;g\in G \}$ and the homogeneous space $G/H=:X$. $X$ allows some geometric structure, and it is natural to denote the state $|{\psi}_{g(x)}\rangle$ corresponding to $x\in X$ by $|x \rangle$ (this is equivalent to considering some local sections $g(\cdot )$ of the homogeneous bundle $G\underset{H}{\to}G/H$):
\begin{align*}
\text{{\small [orbit]}\qquad\quad} & \text{\qquad{\small [states]}} \\
\{ |{\psi}_{g}\rangle ;g\in G \} \text{\quad} & \ni \text{\quad} |{\psi}_{g(x)}\rangle =|x \rangle\\
_{\text{1 to 1}}{\updownarrow} \text{\qquad} & \text{\quad} \text{\qquad} {\updownarrow}_{\text{1 to 1}}\\
G/H=X \text{\quad} & \ni \text{\qquad}x \\
\text{{\small [homogeneous space]}\quad} & \text{\qquad{\small [points]}}
\end{align*}
It is convenient to set $|{\psi}_{0}\rangle=|0\rangle$ as the base point of the manifold $X$. In this way, the set of states $\{ |{\psi}_{g(x)}\rangle=|{\psi}_{g\cdot 0}\rangle=|x\rangle=|x \{g \} \rangle \}$ (in rightmost side of this equation, $g$ means the representative of the equivalence class of $x$) can be identified with a geometric object $X$.

As an additional remark, we obtain the formula of the action $U(g)|{\psi}_{0}\rangle =\exp(i\tilde{\alpha}(g))|{\psi}_{g\cdot 0}\rangle$ for any $g\in G$ with some phase factor $\tilde{\alpha}$ as an extension of $\alpha$ (i.e., $\tilde{\alpha}{\upharpoonright}_{H}=\alpha$). This observation leads us to a viewpoint of bundle structures as shown below:\\
\setlength{\unitlength}{0.7mm}
\begin{picture}(150,40)(-35,-4)
\put(40,0){\makebox(20,10){$x\in G/H$}}
\put(36,20){\makebox(20,10){$g(x)\in G$}}
\put(75,1){\makebox(20,10){$\tilde{M}$}}

\put(78,5){\vector(-1,0){15}}
\put(70,-2){$S^{1}$}
\put(65,20){\vector(3,-2){15}}
\put(72,18){$\pi$}
\put(55,20){\vector(0,-1){10}}
\put(57,13){$H$}
\put(40,10){\vector(0,1){10}}
\put(39,10){\line(1,0){2}}
\put(13,13){\scriptsize{local sections}}
\end{picture}
\\ where $\tilde{M}:=\{ (\exp (i\alpha(g)),x)\in S^{1}\times X \}$. On the basis of the correspondences shown above, a general framework for description of coherent state systems is summarized below:
\begin{itemize}
\item $\{|{\psi}_{g}\rangle ;g\in G \}$ is called a coherent state system associated to $G$ with respect to a unitary irreducible representation $U$ and a vector ${\psi}_{0}\in\mathcal{H}$, and is denoted by $\{G,U,{\psi}_{0} \}$, or $\{G,U,|{\psi}_{0}\rangle \}$.
\item $|{\psi}_{0}\rangle$ and ${\psi}_{0}$ are, respectively, called a standard state and a standard vector of $\{G,U,{\psi}_{0} \}$.
\item $G/H=X$ is called a coherent phase space of $\{G,U,{\psi}_{0} \}$, while $\tilde{M}$ is called a coherent state manifold of $\{G,U,{\psi}_{0} \}$.
\item $|{\psi}_{0}\rangle=|0 \rangle$ as the base point of $X$.
\item $|{\psi}_{g(x)}\rangle=|{\psi}_{g\cdot 0}\rangle=|x\rangle$ for an arbitrary point $x\in X$.
\end{itemize}
There are some general properties of GCS systems \cite{Kla,Per}:
\paragraph{Resolutions of unity on phase spaces}
Since $G$ is a Lie group, $X=G/H$ forms a homogeneous space with a differential structure, and there is a natural action $G\curvearrowright X$. Then we can find a measure $dx$ which is quasi-invariant under this action owing to the group symmetry of $G$. Thus $(X,d\nu)$ is a measurable space, and the following is valid: There exists $C>0$ such that $\int_{X}d\mu (x)|x\rangle\langle x|=\hat{1},\text{ where }d\mu (x):=\frac{1}{C}d\nu (x)$. Apparently, $(X,d\mu)$ is also a measurable space and seen as phase spaces (From now on, we take $d\mu$ rather than $d\nu$ as standard measure of coherent phase spaces). In addition, $X$ is a metric space with a quasi-invariant Riemannian metric $ds^{2}$ which is compatible with $d\mu$ : $(X,d\mu)=:(X,ds^{2})$.
\paragraph{Description for operational quantum physics}
On the basis of the resolution of unity derived above, we can define the corresponding positive operator-valued measure (POVM, in short) on $X$ associated to the coherent state system: Let $\mathcal{B}(X)$ be a Borel set of $X$, then the POVM $M:\mathcal{B}(X)\to\mathcal{L}(\mathcal{H})$ is defined as $M(A):=\int_{x\in A}d\mu(x)|x\rangle\langle x|$ for any $A\in \mathcal{B}(X)$. A triplet $(X,ds^{2},M)$ consisting of a homogeneous space $X=G/H$, a metric $ds^{2}$, and a POVM $M$ (defined above) is also called coherent phase space. It is notable that we can canonically construct the {\em generalized observable} $dM(x):=d\mu(x)|x\rangle\langle x|$.
\paragraph{Expansion on ``CS base" and symbols of states}
Every state $|\psi\rangle\in \mathbb{P}(\mathcal{H})$ can be expanded over coherent state system in the following way: $ |\psi\rangle =\int_{X}d\mu (x)|x\rangle\langle x|\psi\rangle=\int_{X}d\mu (x)\psi (x)|x\rangle $, where $\psi(x):=\langle x|\psi\rangle$ ($\psi :X\to\mathbb{C}$), which is called a symbol of a state $|\psi\rangle$. This symbol satisfies the following formula for inner products: $ \langle \phi|\psi\rangle=\int_{X}d\mu (x)\overline{\phi(x)}\psi(x)$ (These symbols can be naturally seen as $L^{2}$-functions on $X$. $\overline{z}$ denotes the complex conjugate of $z$).
\paragraph{Reproducing kernels}
For arbitrary two points $x,y\in X$, their symbols $\psi(x)$, $\psi(y)$ depend on each other as follows:\\
$\psi(x)=\langle x|\psi\rangle=\int_{y\in X}d\mu(y)\langle x|y\rangle\langle y|\psi\rangle=\int_{y\in X}d\mu(y)K(x,y)\psi(y)$, where $K(x,y):=\langle x|y\rangle$. $K:X\times X\to\mathbb{C}$ is called a reproducing kernel associated to a GCS system $\{G,U,{\psi}_{0} \}$ because it satisfies such property as $K(x,z)=\int_{y\in X}K(x,y)d\mu(y)K(y,z)$. So-called ``overcomplete" linear dependence $|x\rangle=\int_{y\in X}d\mu(y)K(y,x)|y\rangle$ is clearly represented using this reproducing kernel. The factor $\Delta(x,y):=|K(x,y)|^{2}=\overline{\langle x|y\rangle}\langle y|x\rangle=|y(x)|^{2}$ belongs to the space $L^{1}(X)$ because the symbol $y(\cdot)$ is in $L^{2}(X)$. This function plays a central role in our scheme for deriving QPD's as shown in \S\ref{GenQPD}.

\subsection{Semiclassical systems and displacement operators}\label{semicla}

Here we consider the criterion for {\em semiclassical systems}: in other words, ``how to know whether a coherent phase space $X$ has some coordinates which allows physical interpretations". The criterion is, in short, {\em maximality of isotropy subalgebras} of the corresponding Lie algebra defined as below.

Let $\mathfrak{g}$ be the Lie algebra corresponding to the Lie group $G$. Since $G$ is a real group, $\mathfrak{g}$ is a real Lie algebra, and we can construct its complex extension $\mathfrak{g}_{\mathbb{C}}:=\mathfrak{g}{\oplus}_{\mathbb{R}}i\mathfrak{g}$ and a representation $\mathcal{U}$ of $\mathfrak{g}_{\mathbb{C}}$ induced from the unitary irreducible representation $U:G \to GL(\mathcal{H})$ of $G$. Then we can define an algebra $\mathfrak{b}\subset\mathfrak{g}_{\mathbb{C}}$ called isotropy subalgebra with respect to a fixed vector $\psi_{0}\in\mathcal{H}$ as follows: For any $b\in\mathfrak{b}$ there exists $\lambda_{b}\in\mathbb{C}$ such that $\mathcal{U}(b)|\psi_{0}\rangle=\lambda_{b}|\psi_{0}\rangle$. $\mathfrak{b}$ is necessarily a complex subalgebra, and its Hermite conjugate $\overline{\mathfrak{b}}$ is also a subalgebra of $\mathfrak{g}_{\mathbb{C}}$. Fix an arbitrary vector ${\psi}_{0}\in\mathcal{H}$, then a subalgebra $\mathfrak{b}$ of ${\mathfrak{g}}_{\mathbb{C}}$ is called maximal iff $\mathfrak{b}\oplus\overline{\mathfrak{b}}={\mathfrak{g}}_{\mathbb{C}}$, where the direct sum is in the sense of Lie algebras.

Take an isotropy subalgebra $\mathfrak{b}$ with respect to ${\psi}_{0}$ which is maximal in ${\mathfrak{g}}_{\mathbb{C}}$. Then a remarkable proposition holds: The corresponding coherent phase space $X=G/H$, which is a real homogeneous space, is identified with complex homogeneous space $G_{\mathbb{C}}/B$ or $\overline{B}/L$, where $B$, $\overline{B}$, and $L$ are, respectively, the Lie groups corresponding to $\mathfrak{b}$, $\overline{\mathfrak{b}}$, and $\mathfrak{l}$, where $\mathfrak{l}=\mathfrak{b}\cap\overline{\mathfrak{b}}$. It is remarkable that a complex structure is induced in $X$ from $G_{\mathbb{C}}/B$ via the isomorphic relation $X=G/H \simeq G_{\mathbb{C}}/B \simeq \overline{B}/L$ \cite{Borel}.

Such a case can be understood as a typical one that $G$ is a compact semisimple Lie group. Let $T$ be a maximal torus group of $G$ and $\mathfrak{g}$ and $\mathfrak{t}$ be, respectively, the Lie algebra of $G$ and $T$. Let us fix a Cartan base $\{ T_{j},E_{\pm\alpha}\}_{j,\alpha}$ such that $\{T_{j} \}$ spans $T$. We can take the following fundamental geometric objects for representation theory:

$B_{\pm}$: Borel subgroups, $\mathfrak{b}_{\pm}:=Lie(B_{\pm})$; spanned by $\{ T_{j},E_{\pm\alpha}\}_{j,\alpha}$.

$Z_{\pm}$: Nilpotent subgroups, ${\mathfrak{z}}_{\pm}:=Lie(Z_{\pm})$; spanned by $\{E_{\pm\alpha}\}_{\alpha}$.

$T_{\mathbb{C}}$: Complexified group of $T$; $\mathfrak{t}_{\mathbb{C}}:=Lie(T_{\mathbb{C}})$; spanned by $\{ T_{j}\}_{j}$;

$r:=dimT=dim(\mathfrak{t})$: rank of $G$.
\\ Then $X=G/T$ admits a complex structure in a similar way:

\begin{picture}(150,30)(0,3)
\put(46,25){$B_{+}\backslash G_{\mathbb{C}}$}
\put(66,25){$\simeq$}
\put(72,25){$G/T$}
\put(84,25){$\simeq$}
\put(90,25){$G_{\mathbb{C}}/B{-}$}

\put(55,15){$\parallel$}
\put(76,15){$\parallel$}
\put(97,15){$\parallel$}

\put(53,5){$X_{-}$}
\put(75,5){$X$}
\put(95,5){$X_{+}$}
\end{picture}
\\ These homogeneous spaces $X,X_{\pm}$ are called flag manifolds. An essential structure of this isomorphic relation comes from the canonical decomposition in $G_{\mathbb{C}}$, i.e., there are some dense subspace $G_{0}\subset G_{\mathbb{C}}$ which satisfies the following: For any $g\in G_{0}$, there exists ${\xi}_{\pm}\in Z_{\pm},h\in T_{\mathbb{C}},{\eta}_{\pm}\in B_{\pm}$ such that $g={\xi}_{+}h{\xi}_{-}={\eta}_{+}{\xi}_{-}={\xi}_{+}{\eta}_{-}$, and they are unique. In addition, both $X_{\pm}$ admit a Hermitian $G$-invariant metric $ds_{\omega}^{2} = h_{j\overline{k}}d{\xi}_{j}d\overline{{\xi}_{k}}$ and 1-form $\omega = \frac{i}{2}h_{j\overline{k}}d{\xi}_{j}\wedge d\overline{{\xi}_{k}}$ in common  with respect to the corresponding unitary irreducible representation $T$ of $G$. Here $h_{j\overline{k}}:=\frac{\partial}{\partial{\xi}_{j}\partial\overline{{\xi}_{k}}}F(\xi,\overline{\xi})$, and $F$ is the K\"{a}hler potential of $\omega$. The corresponding K\"{a}hler potential $F$ is determined by the Lie-algebraic structure of ${\mathfrak{g}}_{\mathbb{C}}$.

The reason why the K\"{a}hler structure $\omega$ is essential is that any K\"{a}hler manifold can be seen as real symplectic manifold with canonically conjugate variables: Corresponding to the complex coordinates $\xi=(\xi_{j})_{j=1}^{dim_{\mathbb{C}}X}$ of the K\"{a}hler manifold $(X, \omega)$, we can take the symplectic coordinates $(Q;P)=(Q_{j}:=C(\hbar)Re(\xi_{j});P_{j}:=C(\hbar)Im(\xi_{j}))$ ($C(\hbar)$ is some constant and in physical contexts it probably contains the Planck constant $\hbar$ in its formula), and these variables $Q_{j}$,$P_{j}$ play roles as classical observables of the systems. This correspondence allows us to describe the coherent states by the classical variables such as $\{G,U,\psi_{0}\}\ni|\xi\rangle
=|Q;P\rangle
$. This fact leads us to mostly direct correspondence between the generators of $B_{\pm}$ (ladder operators) as q-numbers and the coordinate variables of phase spaces as c-numbers, and is the origin of the word ``semiclassical". From now on, we freely switch these complex and real expression of coherent states as appropriate.

\paragraph{}
Now we introduce a convenient expression of GCS systems with {\em displacement operators} on the basis of the notion of semiclassical systems shown above. We have seen the law of 1-to-1 correspondence between each point of a coherent phase space and a state such as $G/H=X$($=X_{\pm}$)$\ni\xi\leftrightarrow |\xi\rangle$. Considering the viewpoint of bundle structures discussed in \S\ref{GeneralTheory}, for two arbitrary coherent states we can find a particular point of $X=G/H$ which shifts from one to the other, i.e., for all $\xi ,\zeta\in X$, there exists $D:X\to U(\mathcal{H})$ and $\eta\in X$ such that $|\xi\rangle=D(\eta)|\zeta\rangle$. The concrete form of displacement operator $D$ is determined from the algebraic structure of $\mathfrak{g}$ or $\mathfrak{g}^{\mathbb{C}}$ and the representation $\mathcal{U}$ of $\mathfrak{g}$. It is notable that the unitary operators $\{ D(\xi) \}$ do not form a group, but satisfy such a multiplication law as $D(\xi_{1})D(\xi_{2})=D(\xi_{3})\hat{H}(\xi_{1},\xi_{2},\xi_{3})$ (with some $\hat{H}\in U(\mathfrak{h})$). However, this notion of a displacement operator is important because they give another expression of GCS systems:
\[
\{ G,U,\psi_{0} \}=\{ D(\xi)|0\rangle;\xi \in X=G/H \}
\]
This expression of GCS systems involves their (K\"{a}hler) geometric structures in a direct manner. Indeed, for complex-1-dimensional cases (i.e., when $X$ is a Riemannian surface) displacement operators recover symplectic structures of CS phase spaces in the following way:
\begin{equation}
D(\xi)D(\zeta)=\exp [\omega_{X}(\xi,\zeta)]D(\zeta)D(\xi)
\end{equation}
where $\omega_{X}(,)$ is the symplectic form of the K\"{a}hler manifold $(X,\omega )$. This relation is a generalized version of Weyl commutation relation. We can also consider the real expression of displacement operators $|Q;P\rangle=D^{R}(Q;P)|0\rangle$ (Let us omit the index $R$ from now on). Thus, introduction of displacement operators leads us a vivid picture how geometric structures of GCS systems work on state spaces.  For our proposal of formulation of QPD's via geometric structures of GCS systems, these symplectic manifolds should be seen as physical phase spaces, which are nothing but domains of distribution functions defined later.

\section{Coherent state systems and operational quantum physics: Especially for CCR systems}\label{Operational}
In operational quantum physics (especially for open systems), the idea of {\em Wigner distributions} as a class of QPD's tells us some important features of quantum world \cite{geomQM, Wigner}. For instance, it is known that Wigner distributions are useful as degrees of quantum entanglement, whose value is not always positive (This is the origin of the prefix ``quasi"). On the other hand, the {\em Husimi distributions} are also well known, but they are exactly probability distributions which have positive values everywhere \cite{Husimi}. Moreover, there are other QPD's such as the Glauber-Sudarshan distribution \cite{Glau,Suda} and so on, and our goal is some unified viewpoint for these distribution functions. First, let us see that in an example case of CCR algebras in an ordinal way for the purpose of preparing fundamental ideas and notions.

\subsection{Quasi-probability distributions for CCR systems}\label{CCR}
Let $\mathcal{W}_{1}$ be the 1-degree-of-freedom CCR algebra (3-dimensional Heisenberg algebra) generated by $\hat{q}$, $\hat{p}$ and $\hat{I}$ which satisfy the canonical commutation relation $[\hat{q},\hat{p}]=i\hbar\hat{I},[\hat{q},\hat{I}]=[\hat{p},\hat{I}]=0$, or using the creation/annihilation operator, $[\hat{a},\hat{a}^{*}]=\hat{I},[\hat{a},\hat{I}]=[\hat{a}^{*},\hat{I}]=0$ (where $\hat{a}=\frac{1}{\sqrt{2\hbar}}(\hat{q}+i\hat{p}),\hat{a}^{*}=\frac{1}{\sqrt{2\hbar}}(\hat{q}-i\hat{p})$). A conventional description of the GCS system associated to the Lie group $H_{1}={\rm Exp}(\mathcal{W}_{1})$ is written down in the following way: Any $g\in H_{1}$ can be represented as $g=[s;t_{1};t_{2}]$ with the parameters $s,t_{1},t_{2}\in\mathbb{R}$, and let $T_{r}$ be a representation on $\mathcal{H}_{r}$ such that
\[
T_{r}(g)|{\phi}_{0}\rangle=\exp[i(s\hat{I}+t_{1}\hat{a}+t_{2}\hat{a}^{*})]|{\phi}_{0}\rangle
\mbox{\quad for }|{\phi}_{0}\rangle
\in\mathbb{P}(\mathcal{H}_{r}).
\]
Then the arbitrary coherent state in $\{ H_{1},T_{r},|{\phi}_{0}\rangle\}$ is
\begin{align*}
|{\phi}_{g}\rangle &=T_{r}(g)|{\phi}_{0}\rangle =\exp[i(s\hat{I}+t_{1}\hat{a}+t_{2}\hat{a}^{*})]|{\phi}_{0}\rangle\\
&=\exp(\alpha\hat{a}^{*}-\overline{\alpha}\hat{a})\exp(is'\hat{I})|{\phi}_{0}\rangle
\end{align*}
with 2 parameters $\alpha\in\mathbb{C},s'\in\mathbb{R}$ which depend on $s,t_{1},t_{2}$. Since the isotropy subalgebra is $(\hat{I})_{\mathbb{R}}=\{ \exp(is\hat{I})|s\in\mathbb{R}\} \simeq S^{1}$, this GCS system is equivalent to $\{ |{\phi}_{g}\rangle
=D(\alpha )|{\phi}_{0}\rangle \}$ with the standard state $ |{\phi}_{0}\rangle\in\mathbb{P}(\mathcal{H}_{r})$, where $D(\alpha ):=\exp(\alpha\hat{a}^{*}-\overline{\alpha}\hat{a})$, and the arbitrary coherent state is parameterized by $\alpha\in\mathbb{C}\simeq X:=H_{1}/S^{1}$.

More generally, we should take an eigenstate of $\hat{a}$ as the standard state. Then the isotropy subalgebra $\mathcal{B}$ is generated by $\hat{a},\hat{I}$, and $\mathcal{B}= \underset{\mathbb{C}}{span}\{\hat{a},\hat{I}\}$, $\overline{\mathcal{B}}= \underset{\mathbb{C}}{span}\{\hat{a}^{*},\hat{I}\}$ satisfy the maximality condition $\mathcal{B}\oplus\overline{\mathcal{B}}=(\mathcal{W}_{1})_{\mathbb{C}}$. Thus $(H_{1})_{\mathbb{C}}/B\simeq H_{1}/S^{1}\simeq\mathbb{C}=X$ corresponds to a classical phase space $\Gamma=\mathbb{R}^{2}=\{ (q,p)\}$, which is equipped with the K\"{a}hler (symplectic) structure ${\omega}_{H_{1}}=\frac{i}{\pi}d\alpha\wedge d\overline{\alpha}=\frac{1}{\pi\hbar}dq\wedge dp$. The coherent states are also parameterized by $(q,p)\in\Gamma$ : $|\alpha\rangle =\exp(\alpha\hat{a}^{*}-\overline{\alpha}\hat{a})|0\rangle =\exp \left( \frac{i}{\hbar}(p\hat{q}-q\hat{p}) \right)|0\rangle =:|q,p\rangle$.

Instead of formulating QPD's in a direct way, we first define so-called {\em Kernel operators} for CCR CS systems defined as
\begin{equation}\label{eq:CCR}
{\Xi}^{(s)}(z):=\int_{\alpha\in X}\frac{d^{2}\alpha}{\pi}D(\alpha )e^{\frac{s}{2}|\alpha|^{2}-\alpha\overline{z}+\overline{\alpha}z}\mbox{\quad for }z\in\mathbb{C}
\end{equation}
for each $s\in\mathbb{R}$, where $d^{2}\alpha =d({\rm Re}\alpha )d({\rm Im}\alpha )=\frac{1}{2i}d\alpha d\overline{\alpha}$ and $D(\alpha)$ is the coherent displacement operator defined above (see \cite{Brif-Mann}). These kernel operators satisfy the following properties for any $s\in\mathbb{R}$:
\begin{eqnarray*}
& &\mbox{1) }\displaystyle{Tr[{\Xi}^{(s)}(z)]=1} \mbox{\qquad (normalization)},\\
& &\mbox{2) }\displaystyle{\int_{z\in\mathbb{C}}\frac{d^{2}z}{\pi}{\Xi}^{(s)}(z)=1} \mbox{\qquad (completeness)},\\
& &\mbox{3) }\displaystyle{Tr[{\Xi}^{(s)}(z){\Xi}^{(-s)}(z')]={\delta}^{(2)}(z-z')} \mbox{\qquad (orthogonality)}.
\end{eqnarray*}

In the next step we define a family of QPD's as a generalization of well-known Wigner distributions and Husimi distributions as follows: Let $\rho\in\mathcal{S}(\mathcal{H}_{r})$ be a density operator on $\mathcal{H}_{r}$. The functions $\{ F^{(s)}(z)\} _{s\in\mathbb{R}}$ with the parameter $s\in\mathbb{R}$ defined by
\[
F^{(s)}(z):=Tr[\rho{\Xi}^{(-s)}(z)]
\]
is called QPD's of the state $\rho$. Owing to the correspondence of $X=\{ z\in\mathbb{C}\} \simeq{\mathbb{R}}^{2}=\{ (q,p)\}$ ($q={\sqrt{\frac{\hbar}{2}}}(z+\overline{z}),p=-i{\sqrt{\frac{\hbar}{2}}}(z-\overline{z})$), $F^{(s)}(z)$ can be seen as a function $F^{(s)}(q,p)$ of real symplectic variables $(q,p)\in\mathbb{R}^{2}$ (For ease, we use the same notation $F^{(s)}$). Some of the special values of $s$ correspond to the named classes of QPD's:
\begin{eqnarray*}
& &(s=0)\mbox{\quad}F^{(0)}(z)\mbox{: Wigner distribution}\\
& &(s=1)\mbox{\quad}F^{(1)}(z)\mbox{: Husimi distribution}\\
& &(s=-1)\mbox{\quad}F^{(-1)}(z)\mbox{: Glauber-Sudarshan distribution}
\end{eqnarray*}
We can easily check the following properties for the Husimi distribution $F^{(1)}$ and the Glauber-Sudarshan distribution $F^{(-1)}$:
\begin{eqnarray*}
& &\mbox{1) }\displaystyle{F^{(1)}(q,p)=\langle q,p|\rho|q,p\rangle
}\mbox{\quad for any }\rho\in\mathcal{S}(\mathcal{H}),\\
& &\mbox{2) }\displaystyle{\rho =\int\int_{\mathbb{R}^{2}}\frac{dqdp}{2\pi}|q,p\rangle
F^{(-1)}(q,p)\langle q,p|}\mbox{\quad for any }\rho\in\mathcal{S}(\mathcal{H}).
\end{eqnarray*}

Similarly, we can define the {\em symbols of operators} for the CCR system as
\[
F^{(s)}_{A}(z):=Tr[\hat{A}{\Xi}^{(-s)}(z)]
\]
for arbitrary observable $\hat{A}\in\mathcal{W}_{1}$. We also call them QPD's (with respect to $\hat{A}$. $F^{(s)}_{\rho}(z)$ is often denoted by $F^{(s)}(z)$). It is remarkable that the individual mappings of $\{ F^{(s)}_{\cdot}(z):\hat{A}\mapsto\mathbb{C}\}_{z\in\mathbb{C}}$ can be seen as states (in the sense of linear mappings on some appropriate closure of $\mathcal{W}_{1}$) for every fixed point $z$ on the phase space $X=\mathbb{C}=\mathbb{R}^{2}$.

\subsection{Naimark extension and Husimi distributions, Wigner distributions}\label{Naimark}
Another aspects of QPD's (especially Husimi and Wigner distributions) are seen in the contexts of Naimark extension \cite{Ojima89}. The method of Naimark extension shows us a canonical formulation of a spectral measure on a dilated composite system. This construction has a parallelism with quantum measurement schemes $\mathcal{M}\hookrightarrow\mathcal{M}\otimes\mathcal{A}=\mathcal{M}{\rtimes}_{\alpha =id_{\mathcal{M}}}\mathcal{U}(\mathcal{A})\rightarrow\mathcal{M}\otimes\mathcal{A}\twoheadrightarrow\mathcal{M}$ ($\mathcal{M}:\mbox{observed system}, \mathcal{A}:\mbox{probe system}$) \cite{Ojima}. It is essential to consider the coupled system of the observed system $\mathcal{M}$ and the probe $\mathcal{A}$ which have the corresponding Hilbert spaces $\mathcal{H}_{\mathcal{M}}$ and $\mathcal{H}_{\mathcal{A}}$ respectively. Then a generalized observable in $\mathcal{M}$ is reduced to a spectral-decomposable observable on $\mathcal{H}_{\mathcal{M}}\otimes\mathcal{H}_{\mathcal{A}}$. Let us see an example of a scheme of approximately simultaneous measurement of position $q$ and momentum $p$. We can consider a generalized observable $dM(q,p)=|\psi_{q,p}\rangle
\langle \psi_{q,p}|\displaystyle{\frac{dqdp}{2\pi}}$ and calculate the expectation value for the state $\rho$ of its Fourier transform as
\begin{equation*}
Tr \left[ \rho\int e^{i(uq+vp)}dM(q,p) \right] =Tr[\rho\otimes |\overline{\psi}\rangle
\langle \overline{\psi}| e^{i(u\hat{q}_{comp}+v\hat{p}_{comp})}],
\end{equation*}
where $\langle q|\overline{\psi}\rangle
:=\overline{\langle q|\psi\rangle
}=\overline{\psi}(q)$, $\hat{q}_{comp}:=\hat{q}\otimes\hat{1}-\hat{1}\otimes\hat{q}$, $\hat{p}_{comp}:=\hat{p}\otimes\hat{1}+\hat{1}\otimes\hat{p}$ \cite{Ojima89}. Since $[\hat{q}_{comp},\hat{p}_{comp}]=[\hat{q},\hat{p}]\otimes\hat{1}-\hat{1}\otimes[\hat{q},\hat{p}]=0$, $\hat{q}_{comp}$ and $\hat{p}_{comp}$ are simultaneously diagonalizable and with their spectral measure $dN(q,p)$ and we obtain the decomposition
\[
e^{i(u\hat{q}_{comp}+v\hat{p}_{comp})}=\int e^{i(uq+vp)}dN(q,p).
\]
Then the followings are valid \cite{Ojima89}:
\begin{proposition}
\begin{eqnarray}
\int e^{i(uq+vp)}Tr[\rho dM(q,p)] &=& \int e^{i(uq+vp)}Tr[\rho\otimes |\overline{\psi}\rangle
\langle \overline{\psi}| dN(q,p)], \nonumber\\
Tr[\rho dM(q,p)] &=& Tr[\rho\otimes |\overline{\psi}\rangle
\langle \overline{\psi}| dN(q,p)]\\
&=& \frac{1}{2\pi}\langle \psi_{q,p}|\rho|\psi_{q,p}\rangle
dqdp=:\rho(q,p)dqdp, \nonumber
\end{eqnarray}
where $\rho\in\mathcal{S}(\mathcal{H}_{\mathcal{M}})$ and $|\psi\rangle
\in\mathbb{P}(\mathcal{H}_{\mathcal{A}})$, and $dM$, $dN$ are defined above. Suppose $|\psi\rangle =|\psi_{0}\rangle :=|\alpha =0\rangle \in \{H_{1},T_{r},\phi_{0} \}$, then the simultaneous probability distribution $\rho(q,p)$ is nothing but the Husimi distribution.
\end{proposition}
Note that $|\psi_{0}\rangle$ is a Gaussian state defined by $\langle q|\psi_{0}\rangle=\frac{1}{(\pi d)^{1/4}}e^{-\frac{q^{2}}{2d}}$ with the variance $d(>0)$, which is the minimizer of the uncertainty relation. 

Another discussion from the viewpoint of characteristic functions can be done as follows: $Z_{\rho}(u,v):=Tr[\rho e^{i(u\hat{q}+v\hat{p})}]$ (: the quantum Fourier transform of $\rho$) is called the quantum characteristic function, which is corresponding to the (classical) characteristic function as the Fourier transform of the probability distribution $\rho(q,p)$. Then the Wigner distribution is given in the following way:
\[
F^{(0)}_{\rho}(q,p):=\int\frac{dudv}{(2\pi)^{2}}e^{-i(uq+vp)}Z_{\rho}(u,v)=\mathcal{F}^{-1}[Z_{\rho}](q,p).
\]
Because of the non-commutativity of $\hat{q}$ and $\hat{p}$, $F^{(0)}_{\rho}$ is not always positive. The probability distribution $\rho(q,p)$ (with positive values) is obtained via the construction of the coupling of the observed $\rho$ and the probe $|\overline{\psi}\rangle
\langle \overline{\psi}|$ as shown below \cite{Ojima89}:
\begin{eqnarray}
\rho(q,p) &=& \int\frac{dudv}{(2\pi)^{2}}e^{-i(uq+vp)}Tr[\rho e^{i(u\hat{q}+v\hat{p})}]\langle \psi|e^{-i(u\hat{q}+v\hat{p})}|\psi\rangle
 \nonumber\\
&=& \int dq'dp'F^{(0)}_{\rho}(q',p')F^{(0)}_{|\overline{\psi}\rangle
\langle \overline{\psi}|}(q'-q,p'-p)\\
&=& \frac{1}{dqdp}Tr[\rho\otimes|\overline{\psi}\rangle
\langle \overline{\psi}|dN(q,p)]. \nonumber
\end{eqnarray}
The Husimi distribution $F_{\rho}^{(-1)}$ is the particular case for the choice of $|\psi\rangle
=|\psi_{0}\rangle
$: $F_{\rho}^{(-1)}(q,p)=\displaystyle{\frac{1}{dqdp}}Tr[\rho\otimes|0\rangle
\langle 0|dN(q,p)]$.

We can find similar structures on arbitrary systems with Lie-group symmetries, which are associated with their GCS systems $\{ G,U,|0\rangle
\}$ ($U:G\to\mathcal{L}(\mathcal{H})$) with the phase space $(X,d\mu)$. From the general theory in \S\ref{GeneralTheory}, we can pick up some set of quantum observables corresponding to appropriate classical symplectic observables according to the criterion of semiclassical condition in \S\ref{semicla}. The essence is the correspondences between the coherent states and points on the phase space via the orbit method \cite{Kirillov,Kirillov2}, shown in the following diagram:\\
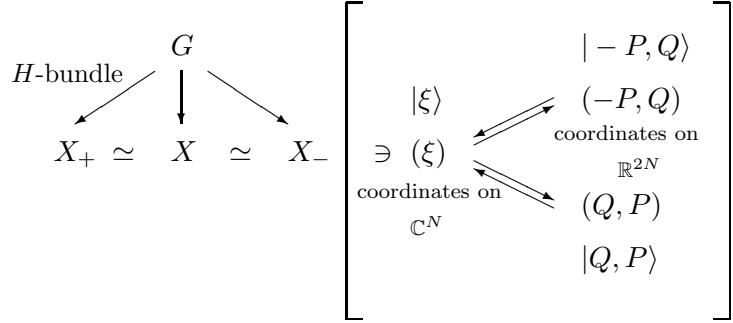
\begin{picture}(150,72)(0,-35)
\put(55,0){$X$}
\put(55,20){$G$}
\put(66,0){$\simeq$}
\put(44,0){$\simeq$}
\put(77,0){$X_{-}$}
\put(33,0){$X_{+}$}

\put(93,0){$\ni$}
\put(100,0){$(\xi)$}
\put(132,-10){$(Q,P)$}
\put(132,10){$(-P,Q)$}
\put(100,10){$|\xi\rangle$}
\put(132,-20){$|Q,P\rangle$}
\put(132,20){$|-P,Q\rangle$}

\put(90,-7){\scriptsize{coordinates on}} 
\put(100,-14){\scriptsize{$\mathbb{C}^{N}$}}
\put(127,4){\scriptsize{coordinates on}}
\put(139,-3){\scriptsize{$\mathbb{R}^{2N}$}}

\put(88,-30){\line(0,1){60}}
\put(160,-30){\line(0,1){60}}
\put(88,30){\line(1,0){3}}
\put(88,-30){\line(1,0){3}}
\put(160,30){\line(-1,0){3}}
\put(160,-30){\line(-1,0){3}}

\put(62,17){\vector(3,-2){15}}
\put(52,17){\vector(-3,-2){15}}
\put(57,17){\vector(0,-1){10}}
\put(25,15){\small{$H$-bundle}}
\put(112,0){\vector(2,-1){15}}
\put(127,-9){\vector(-2,1){15}}
\put(112,3){\vector(2,1){15}}
\put(127,12){\vector(-2,-1){15}}
\end{picture}
\\ where $(Q;P)=(Q_{1},\cdots ,Q_{N};P_{1},\cdots ,P_{N})$ and $\xi_{j}:=Q_{j}+iP_{j}$ denote the symplectic variables of the phase space $X$ ($N=dim_{\mathbb{C}}X$) and the corresponding complex variables respectively. A choice of some local section of $G\to X$ is equivalent to consideration of such particular symplectic coordinates. The generalized observables with coherent states are written down as
\begin{eqnarray*}
dM(Q;P) &=& d\mu(Q;P)|Q;P\rangle\langle Q;P|\\
&=& d\mu(\xi_{1},\cdots ,\xi_{N})|\xi_{1},\cdots ,\xi_{N}\rangle
\langle \xi_{1},\cdots ,\xi_{N}|.
\end{eqnarray*}
For them, we can obtain the probability distribution of approximately simultaneous measurement
\begin{equation}
\rho(Q;P)d\mu(Q;P)=Tr[\rho\otimes|Q=0;P=0\rangle
\langle Q=0;P=0|dN(Q;P)]\label{Husimi1}
\end{equation}
as well, where $dN(Q;P)=\mathcal{F}[e^{i(u\cdot\hat{Q}_{comp}+v\cdot\hat{P}_{comp})}]$, $\hat{Q}_{comp}:=\hat{Q}\otimes\hat{1}-\hat{1}\otimes\hat{Q}$, $\hat{P}_{comp}:=\hat{P}\otimes\hat{1}+\hat{1}\otimes\hat{P}$. This probability distribution is equivalent to the situation of minimum uncertainty because the action of the displacement operator $D$ on the state space $\mathbb{P}(\mathcal{H})$ preserves the variances of the quadratic Casimir operator $C_{2}:=\sum_{j=1}^{r}\hat{T_{j}}^2+\sum_{\alpha\in R_{+}}(\hat{E}_{\alpha}\hat{E}_{-\alpha}+\hat{E}_{-\alpha}\hat{E}_{\alpha})$ ($\{ \hat{T}_{j},\hat{E}_{\alpha}\}$: the Cartan base, $R_{+}$: the positive roots) : $(\Delta C_{2})^{2}_{\xi}=\langle \xi|\hat{C_{2}}^{2}|\xi\rangle-\langle \xi|\hat{C_{2}}|\xi\rangle
^{2}=(\Delta C_{2})^{2}_{\xi =0}=min(\Delta C_{2})^{2}$. The probe state $|Q=0;P=0\rangle$ indicates a neutral position of the measurement (More detailed idea and roles of neutral position are seen in \cite{H-O,Ojima}). From this viewpoint, it is natural to generalize the idea of identifying Husimi distributions as least uncertain calibration of probes for approximately simultaneous measurement schemes:
\begin{definition}\label{Husimi2}
We define the generalized Husimi distribution for $\rho\in\mathcal{S}(\mathcal{H})$ as $F^{(1)}_{\rho}(Q;P):=\langle Q;P|\rho|Q;P\rangle=\langle \xi|\rho|\xi\rangle$, and similarly, for an arbitrary observable $\hat{A}\in\mathfrak{g}^{\prime\prime}=\overline{\mathfrak{g}}^{weak}$ as $F^{(1)}_{A}(Q;P):=\langle Q;P|\hat{A}|Q;P\rangle=\langle \xi|\hat{A}|\xi\rangle$.
\end{definition}
As an additional remark, this representation of operators $(\hat{A}\mapsto\langle \xi|\hat{A}|\xi\rangle)$ is recognizable as a GNS representation $\{\mathcal{H},\mathcal{U},|\xi\rangle
\}$.

From the discussion in this section, it is natural to consider that eq.(\ref{Husimi1}) or Def.\ref{Husimi2} as a natural generalization of Husimi distributions which contains well-known cases such as the Heisenberg group or $SU(2)$ group. As our second important result, we introduce the formula of generalized QPD's in the next section with use of a scheme of deforming generalized Husimi distributions in Def.\ref{Husimi2}.

\section{Generalized quasi-probability distributions with coherent state systems}\label{GenQPD}
So far we have surveyed general aspects of GCS systems and how they are related to the contexts of operational quantum physics. In this section, let us start with the case where we could have taken some coherent phase space with real symplectic forms successfully, and construct families of distribution functions on these spaces as generalized QPD's. There are some works for construction of Wigner and Husimi distributions in general cases, but our method has rather geometric and harmonic-analytic aspects and which can be said as the generalization of the method by V\'{a}rilly-Gracia-Band\'{i}a \cite{V-G}. Our scheme shows how to formulate QPD's more directly from the representation-theoretical ingredients of GCS systems.

In this section, we always denote coordinates of phase spaces by complex variables on the basis of correspondence mentioned in \S\ref{Naimark}.

\subsection{Fourier duality and kernel operators}\label{FD}
As a preparation for discussion on QPD's, let us start with Fourier analyses on general Lie groups. Since the general theory of Fourier duality is too complexified, here we have a short review to pick up some essences which are useful for us. Let $G$ be a Lie group and $\mathfrak{g}$ be the corresponding Lie algebra and take their representations in the similar way we have done in \S\ref{GeneralTheory}. For ease of discussion, let us focus the case with some fundamental series of representations $\{ U^{j}\}$, such as ladder representations of $SU(2)$ or $SU(1,1)$, and so on. Then, for each $U^{j}$, we can determine the coherent phase space $X$ with respect to $\{ G,U^{j},\psi_{0}\}$, or representation theoretically, one of the components of the dual object $(\hat{G})_{j}\subset\hat{G}=\{ (j,\xi);j\in\mathbb{N},\xi\in X\}$. More generally, our discussion here is ensured owing to the orbit method \cite{Kirillov,Kirillov2}, which shows that the orbit of the coadjoint action $G\curvearrowright\mathfrak{g}^{*}$ carries a natural symplectic structure and can be seen as an abstract phase space.

In the next step, we define the Fourier kernel on $G\times\hat{G}$ according to the scheme of V\'{a}rilly and Gracia-Bond\'{i}a \cite{V-G};
\begin{eqnarray*}
W:G\times\hat{G}\to\mathbb{C}\\
W(g;j,\xi):=Tr[U^{j}(g)\Xi^{j}(\xi)]
\end{eqnarray*}
where $\{ \Xi^{j} \}$ are Stratonovich-Weyl (SW, in short) kernel operators satisfying the following properties:
\begin{align*}
& \mbox{\rm (K.1) }\Xi^{j}(\xi )=\Xi^{j}(\xi )^{*} \mbox{\quad for }\forall\xi\in X \mbox{\quad (self-adjointness)},\\
& \mbox{\rm (K.2) }\Xi^{j}(g\cdot\xi )=U^{j}(g)\Xi^{j}(\xi )U^{j}(g^{-1}) \mbox{\quad for }\forall\xi\in X,\forall g\in G\\
& \mbox{ \qquad (covariance with $G\curvearrowright X$)},\\
& \mbox{\rm (K.3) }Tr[\Xi^{j}(\xi )]=1 \mbox{\quad (normalization as trace-class operators)},\\
& \mbox{\rm (K.4) }\int_{X}d\mu (\xi )\Xi^{j}(\xi )=1 \mbox{\quad (normalization as QPD's)},\\
& \mbox{\rm (K.5) }Tr[\Xi^{j}(\xi )\Xi^{j}(\xi ')]=C_{5}{\delta}^{(N)}(\xi -\xi ') \mbox{\quad (orthogonality relation)}.
\end{align*}
where $C_{5}$ is some positive constant which may vary according to conventions; we set $C_{5}=1$ hereafter. Then we can define the Fourier transform on $G$ with $W(g;j,\xi)$ and the invariant measure $dg$ as follows:
\[
(\mathcal{F}f)(j,\xi):=\int_{G}W(g;j,\xi)f(g)dg \mbox{\quad for } f\in L^{\circ}(G)
\]
(where $L^{\circ}(G)$ is some appropriate function space like $L^{1}(G)\cap L^{2}(G)$ or $L^{\infty}(G)\cap L^{2}(G)$, or others). It is an important fact in our context that $W(g;j,\xi)$ can be recognized as a function on the coherent phase spaces $\{X=X^{j}\}$ (each $X^{j}$ is derived from $\{G,U^{j},\psi_{0}\}$) by the identification of $W(g;j,\xi)=:[W(g)]^{j}(\xi)$. Thus, the dual object $[W(g)]^{j}$ of $g\in G$ with respect to the representation $U^{j}$ gives a distribution function on the phase space $X^{j}$. Our guiding principle is to formulate some family of such functions on the basis of the theory of GCS systems.

To clarify the correspondence to Wigner or Husimi distributions, let us rewrite the axioms of SW kernels to the words of 1-to-1 linear mappings $(\mathcal{M}\ni\hat{A}\mapsto W_{A})$ ($\mathcal{M}$ is the v.N. algebra describing the observed system acted by the group $G$) as shown below (From now on, we suppose that some particular representation $U=U^{(j)}$ is fixed):
\begin{align*}
& \mbox{\rm (S.1) }W_{A^{*}}(\xi )=\overline{W_{A}(\xi )} \mbox{\quad for }\forall A\in\mathcal{M},\forall\xi\in X \mbox{\quad (self-adjointness),}\\
& \mbox{\rm (S.2) }W_{g\cdot A}(\xi )=W_{A}(g\cdot\xi ) \mbox{\quad where }g\cdot A=U(g)^{*}AU(g)\\
& \mbox{\qquad\qquad for }\forall A\in\mathcal{M},\forall\xi\in X,\forall g\in G \mbox{\quad (covariance with $G\curvearrowright X$),}\\
& \mbox{\rm (S.3) }\int_{X}d\mu (\xi )W_{A}(\xi )=Tr(A) \mbox{\quad for }\forall A\in\mathcal{M} \mbox{\quad (normalization as QPD's),}\\
& \mbox{\rm (S.4) }\int_{X}d\mu (\xi )W_{A}(\xi )W_{B}(\xi )=Tr(AB) \mbox{\quad for }\forall A,B\in\mathcal{M}.
\end{align*}
Indeed, the Wigner distributions obtained in \S\ref{CCR} satisfy these conditions, but this does not necessarily hold true for the Husimi distributions and the Glauber-Sudarshan distributions. However, the following equation is valid:
\[
\int_{X}d\mu (\xi )F^{(1)}_{A}(\xi )F^{(-1)}_{B}(\xi )=\int_{X}d\mu (\xi )F^{(-1)}_{A}(\xi )F^{(1)}_{B}(\xi )=Tr(AB).
\]
Thus it is natural to claim
\begin{align*}
& \mbox{\rm (S.4')} \mbox{\qquad\qquad}\int_{X}d\mu (\xi )F^{(s)}_{A}(\xi )F^{(-s)}_{B}(\xi )=Tr(AB)\mbox{\qquad\qquad}
\end{align*}
for generalized QPD's, or equivalently,
\begin{align*}
& \mbox{\rm (K.5')} \mbox{\qquad\qquad}Tr[\Xi^{(-s)}(\xi )\Xi^{(s)}(\xi ')]={\delta}^{(N)}(\xi -\xi ')\mbox{\qquad\qquad}
\end{align*}
for kernel operators (for (S.1)-(S.3) or (K.1)-(K.4), $W$ or $\Xi^{j}$ are simply replaced by $F^{(s)}$ or $\Xi^{(-s)}$, respectively). The axioms (K.1)-(K.4) and (K.5') are equivalent to what Brif and Mann postulated for QPD's on homogeneous spaces in \cite{Brif-Mann}. Each component of a 1-parameter family of operators $\{ \Xi^{(-s)}(\xi){\}}_{s\in\mathbb{R}}$ is called a kernel operator associated to $\{ G,U,\psi_{0}\}$ if and only if they satisfy (K.1)-(K.4) and (K.5'), and the symbols of operators $\{ {\Sigma}^{(s)}:\mathcal{L}(\mathcal{H})\to Map(X,\mathbb{C}) \} _{s\in\mathbb{R}}$ defined by ${\Sigma}^{(s)}(\hat{A}):=Tr[\hat{A}\Xi^{(-s)}(\xi )]$ can be considered as well. For every fixed $\xi\in X$, they give the (normal) states ${\Sigma}^{(s)}:\mathcal{L}(\mathcal{H})\to\mathbb{C}$ of the system. Taking density operators $\rho\in\mathcal{S}(\mathcal{H})$, we can obtain the QPD's in ordinary meaning as $F^{(s)}_{\rho}(\xi ):={\Sigma}_{\xi}^{(s)}(\rho)=Tr[\rho \Xi^{(-s)}(\xi )]$. It is remarkable that QPD's (for states) and kernel operators are mutually Fourier dual via the density operator $\rho$.

\subsection{Formulation of QPD's from GCS systems}\label{FinalProp}
The remaining problem is how to construct concrete formulae of QPD's on the basis of our discussion related to operational or harmonic-analytic contexts. We have already found the rationale for identifying Husimi distributions as eq.(\ref{Husimi1}) or Def.\ref{Husimi2}, therefore our strategy is as follows: We identify the Husimi distribution as $F^{(1)}_{A}(\xi )=Tr[\hat{A}\Xi^{(-1)}(\xi )]=\langle \xi|\hat{A}|\xi\rangle
$, and we deform this formula to obtain QPD's $\{ F^{(s)}\}_{s\in\mathbb{R}}$. First, we find a possible formula of distribution $F^{(-1)}_{A}$ as the dual of $F^{(1)}_{A}$. For this purpose, we suppose the expansion formula of $\hat{A}$ as $\hat{A}=:\int_{X}d\mu(\xi)P_{A}(\xi)|\xi\rangle
\langle \xi|$, and let us calculate $\int_{X}d\mu(\xi)F^{(1)}_{A}(\xi)P_{B}(\xi)$:
\begin{align*}
&\int_{X}d\mu(\xi)\langle \xi|\hat{A}|\xi\rangle P_{B}(\xi) = Tr \left[ \int_{X}d\mu(\xi)\hat{A}P_{B}(\xi)|\xi\rangle\langle \xi| \right]\\
&= Tr \left[ \hat{A}\int_{X}d\mu(\xi)P_{B}(\xi)|\xi\rangle\langle \xi| \right]
= Tr(\hat{A}\hat{B})= Tr(\hat{B}\hat{A})\\
&= \int_{X}d\mu(\xi)\langle \xi|\hat{B}|\xi\rangle P_{A}(\xi)
= \int_{X}d\mu(\xi)P_{A}(\xi)\langle \xi|\hat{B}|\xi\rangle .
\end{align*}
From this calculation, we obtain the relation
\[
\int_{X}d\mu(\xi)F^{(1)}_{A}(\xi)P_{B}(\xi)=\int_{X}d\mu(\xi)P_{A}(\xi)F^{(1)}_{B}(\xi)=Tr(\hat{A}\hat{B}),
\]
so it is natural to claim $F^{(-1)}_{A}=P_{A}$. Second, we can obtain the relation between $F^{(1)}_{A}$ and $P_{A}$ with the reproducing kernel $K$ with respect to $\{ G,U,\psi_{0}\}$ as follows:
\begin{align*}
F^{(1)}_{A}(\xi) &= \langle \xi|\hat{A}|\xi\rangle=\int_{X}d\mu(\eta)P_{A}(\eta)\left| \langle \xi|\eta\rangle\right| ^{2}\\
&= \int_{X}d\mu(\eta)P_{A}(\eta)\left| K(\xi,\eta)\right| ^{2}\\
\mbox{{\Huge[}{\quad}} &\overset{\mbox{{\scriptsize to be}}}{=} \int_{X}d\mu(\eta)F^{(-1)}_{A}(\eta)\left| K(\xi,\eta)\right| ^{2}\mbox{\quad\Huge{]}}.
\end{align*}
In order to solve the last equation for $F^{(-1)}_{A}(\eta)$, we introduce exponentials $\Delta^{s}$ as $\Delta^{s}(\xi,\eta):=\sum_{j}\alpha_{j}^{s}\overline{\phi_{j}(\xi)}\phi_{j}(\eta)$ via an expansion formula $\Delta(\xi,\eta)=\sum_{j}\alpha_{j}\overline{\phi_{j}(\xi)}\phi_{j}(\eta)=:\sum_{j}\alpha_{j}v_{j}(\xi,\eta)$ over $L^{2}$-CONS (complete orthogonal normal system consisting of $L^{2}$-functions) $\{ \phi_{j}\}$ with the orthogonal relations $\int_{X}d\mu(\eta)v_{j}(\xi,\eta)v_{j}(\eta,\zeta)=\delta_{j,k}v_{k}(\xi,\zeta)$. This definition is consistent with the convolution-type product $\Delta^{s}*\Delta^{t}:=\int_{X}d\mu(\eta)\Delta^{s}(\xi,\eta)\Delta^{t}(\eta,\zeta)=\Delta^{s+t}(\xi,\zeta)$, and such basis can be always constructed by Gram-Schmidt method via the orthogonal relations. Then we can obtain
\begin{equation*}
F^{(-1)}_{A}(\xi)=\displaystyle{\int_{X}}d\mu(\eta)F^{(1)}_{A}(\eta)\Delta^{-1}(\xi,\eta),
\end{equation*}
and a naturally generalized formula of QPD's with $\Delta(\xi,\eta):=|K(\xi,\eta)|^{2}$:
\begin{eqnarray}\label{GenSol}
F^{(s)}_{A}(\xi)&:=&\displaystyle{\int_{X}}d\mu(\eta)F^{(t)}_{A}(\eta)\Delta^{\frac{s-t}{2}}(\xi,\eta)
=\displaystyle{\int_{X}}d\mu(\eta)F^{(1)}_{A}(\eta)\Delta^{\frac{s-1}{2}}(\xi,\eta)\nonumber\\
&=&\displaystyle{\int_{X}}d\mu(\eta)\langle \eta|\hat{A}|\eta\rangle\Delta^{\frac{s-1}{2}}(\xi,\eta).
\end{eqnarray}
We can also write down the formula of SW kernels via quantum Fourier transform of eq.(\ref{GenSol}):
\begin{eqnarray}\label{GenSol2}
\Xi^{(-s)}(\xi)&=&\displaystyle{\int_{X}}d\mu(\eta)\Xi^{(-t)}(\eta)\Delta^{\frac{s-t}{2}}(\xi,\eta)=\displaystyle{\int_{X}}d\mu(\eta)|\eta\rangle\langle \eta|\Delta^{\frac{s-1}{2}}(\xi,\eta),\nonumber\\
\Xi^{(-1)}(\xi)&=&|\xi\rangle\langle \xi|.
\end{eqnarray}
Let us take the $L^{2}$-CONS $\{ Y_{j}\}$ which consists of the harmonic functions on $X$ and write down the expansion formula $\Delta(\xi,\eta)=\sum_{j}\upsilon_{j}\overline{Y_{j}(\xi)}Y_{j}(\eta)$. Then we obtain
\begin{eqnarray*}
\Xi^{(s^{\prime})}(\xi)&=&\displaystyle{\int_{X}}d\mu(\eta)\Xi^{(-1)}(\eta)\sum_{j}\upsilon_{j}^{-\frac{s^{\prime}+1}{2}}\overline{Y_{j}(\xi)}Y_{j}(\eta)\\
&=&\displaystyle{\sum_{j}}\upsilon_{j}^{-\frac{s^{\prime}}{2}}\overline{Y_{j}(\xi)}\left( \upsilon_{j}^{-\frac{1}{2}}\displaystyle{\int_{X}}d\mu(\eta)Y_{j}(\eta)|\eta\rangle\langle \eta|\right).
\end{eqnarray*}
This formula of SW kernels is equivalent to the one given by Brif and Mann in \cite{Brif-Mann}. Since the axioms (K.1)-(K.4) and (K.5') are constructed with quantities which is invariant under exchanges of CON basis, our formulae (\ref{GenSol2}) also satisfy them. Accordingly, we have proved our second claim in this paper:
\begin{proposition}\label{GenSol3}
$\Xi^{(-s)}$ in {\em eq.(\ref{GenSol2})} satisfies {\em (K.1)-(K.4)} and {\em (K.5')} and $F^{(s)}$ in {\em eq.(\ref{GenSol})} satisfies {\em (S.1)-(S.3)} and {\em (S.4')}, so they well define QPD's (Husimi distributions for $s=1$ and other QPD's for $s\neq 1$) associated to GCS systems:
\begin{eqnarray*}
&\hat{A}&\mapsto\langle \xi|\hat{A}|\xi\rangle,\\
&\hat{A}&\mapsto Tr\left[ \hat{A}\Xi^{(-s)}(\xi)\right]= \displaystyle{\int_{X}}d\mu(\eta)\langle \eta|\hat{A}\Delta^{\frac{s-1}{2}}(\xi,\eta)|\eta\rangle\mbox{\quad}(s\neq 1).
\end{eqnarray*}
Especially, for the case $s=0$ and $s=-1$, this scheme gives Wigner distributions and Glauber-Sudarshan distributions respectively.
\end{proposition}

Moreover, we can clarify the relation between {\em weak measurements} and our scheme as follows: Let us expand $\Delta^{\frac{s-1}{2}}(\xi,\eta)=\sum_{j}\alpha_{j}^{\frac{s-1}{2}}\overline{\phi_{j}(\xi)}\phi_{j}(\eta)$ with an arbitrary $L^2$-CONS $\{\phi_{j}\}$. Then the QPD's in the case of $s\neq 1$ in eq.{\rm (\ref{GenSol3})} can be represented with the weak value $W_{\xi,\eta}(\hat{A}):=\displaystyle{\frac{\langle\eta|\hat{A}|\xi\rangle}{\langle\eta|\xi\rangle}}$ of $\hat{A}$ with respect to the pre- and post-selected state $|\xi\rangle $ and $|\eta\rangle $ \cite{Aha,Aha2} as follows:
\begin{eqnarray*}
\hat{A}\mapsto&\displaystyle{\int_{X}}d\mu(\eta)\langle\eta|\hat{A}\sum_{j}\alpha_{j}^{\frac{s-1}{2}}\overline{\phi_{j}(\xi)}\phi_{j}(\eta)|\eta\rangle\\
=&\displaystyle{\int_{X}}d\mu(\eta)\langle\eta|\hat{A}\sum_{j}\alpha_{j}^{\frac{s-1}{2}}\langle\phi_{j}|\xi\rangle\langle\eta|\phi_{j}\rangle|\eta\rangle\\
=&\displaystyle{\sum_{j}}\alpha_{j}^{\frac{s-1}{2}}\langle\phi_{j}|\left( {\int_{X}}d\mu(\eta)\langle\eta|\hat{A}|\xi\rangle\langle\eta|\eta\rangle \right)|\phi_{j}\rangle\\
=&\displaystyle{\sum_{j}}\alpha_{j}^{\frac{s-1}{2}}\langle\phi_{j}|{\int_{X}}d\mu(\eta)W_{\xi,\eta}(\hat{A})K(\xi,\eta)|\phi_{j}\rangle.
\end{eqnarray*}
This result means that QPD's for $s\neq 1$ can be seen as integrated observed data of approximately simultaneous measurements via weak values. For $s>1$, the factor $\Delta^{\frac{s-1}{2}}$ is understood as a mollifier and in such cases coarse-grained processes attends the measurement schemes. For $s<1$, in contrast, the factor causes anti-mollifying effect and related to recovering of information of quantum systems. In fact, the Wigner distributions $F^{(0)}$ can be recognized as characteristic quantities of quantum entanglement, and on the other hand, the Husimi distributions $F^{(1)}$ are the most appropriate for classical picture. From our discussion, such mutual relations can be found all over the QPD's $\{ F^{(s)};s\in\mathbb{R}\}$ via losing/recovering of information of quantum systems.\\
\setlength{\unitlength}{0.7mm}
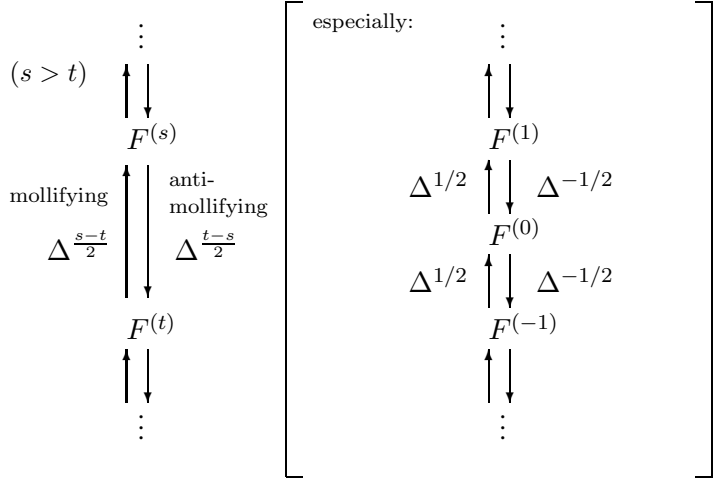
\begin{picture}(150,95)(-55,-18)

\put(25,70){\mbox{{\scriptsize especially:}}}
\put(58,-1){\vector(0,1){10}}
\put(58,17){\vector(0,1){10}}
\put(58,35){\vector(0,1){10}}
\put(58,53){\vector(0,1){10}}
\put(62,9){\vector(0,-1){10}}
\put(62,27){\vector(0,-1){10}}
\put(62,45){\vector(0,-1){10}}
\put(62,63){\vector(0,-1){10}}
\put(60,-9){$\cdot$}
\put(60,-7){$\cdot$}
\put(60,-5){$\cdot$}
\put(60,65){$\cdot$}
\put(60,67){$\cdot$}
\put(60,69){$\cdot$}
\put(58,29){$F^{(0)}$}
\put(58,47){$F^{(1)}$}
\put(58,11){$F^{(-1)}$}
\put(43,20){$\Delta^{1/2}$}
\put(43,38){$\Delta^{1/2}$}
\put(67,20){$\Delta^{-1/2}$}
\put(67,38){$\Delta^{-1/2}$}

\put(20,-15){\line(0,1){90}}
\put(20,-15){\line(1,0){3}}
\put(20,75){\line(1,0){3}}
\put(100,-15){\line(0,1){90}}
\put(100,-15){\line(-1,0){3}}
\put(100,75){\line(-1,0){3}}

\put(-32,60){\mbox{{\small ($s>t$)}}}
\put(-32,37){\mbox{{\scriptsize mollifying}}}
\put(-2,40){\mbox{{\scriptsize anti-}}}
\put(-2,35){\mbox{{\scriptsize mollifying}}}
\put(-10,-1){\vector(0,1){10}}
\put(-10,19){\vector(0,1){25}}
\put(-10,53){\vector(0,1){10}}
\put(-6,9){\vector(0,-1){10}}
\put(-6,44){\vector(0,-1){25}}
\put(-6,63){\vector(0,-1){10}}
\put(-8,-9){$\cdot$}
\put(-8,-7){$\cdot$}
\put(-8,-5){$\cdot$}
\put(-8,65){$\cdot$}
\put(-8,67){$\cdot$}
\put(-8,69){$\cdot$}
\put(-10,47){$F^{(s)}$}
\put(-10,11){$F^{(t)}$}
\put(-25,26){$\Delta^{\frac{s-t}{2}}$}
\put(-1,26){$\Delta^{\frac{t-s}{2}}$}
\end{picture}

Now we have obtained a canonical scheme of formulating QPD's $\{ F^{(s)}\}$ via GCS systems for each $s\in\mathbb{R}$, and have understood that two QPD systems $\{ F^{(s)}\}$ and $\{ F^{(t)}\}$ ($s\neq t$) are mutually related as eq.(\ref{GenSol}) or eq.(\ref{GenSol2}). As shown in \S\ref{FD}, QPD's play roles as Fourier transforms of group elements, or corresponding operators in quantum systems. So we can transform eq.(\ref{DynEq}) shown in the beginning of this paper to equations on a coherent phase space $X$ as follows: The formula is, again,
\[
\frac{\partial\rho}{\partial \tau}=\mathcal{P}(\rho,\hat{A_{1}},\hat{A_{2}},\cdots)\mbox{\qquad}(\tau\in\mathbb{R}).
\]
Since $\mathcal{P}$ is a polynomial, $\mathcal{P}(\rho,\hat{A_{1}},\hat{A_{2}},
\cdots)$ is also a element of the *-algebra under consideration, so there exists the QPD $F^{(s)}_{\mathcal{P}}$ of the operator $\hat{\mathcal{P}}$. If we can find some differential operator $D_{s,\mathcal{P}}(\xi,\overline{\xi},\partial_{\xi},\partial_{\overline{\xi}})$ such that $F^{(s)}_{\mathcal{P}}=D_{s,\mathcal{P}}F^{(s)}_{\rho}$, we obtain the formula of dynamical equations of functions on $X$ as
\[
\frac{\partial}{\partial \tau}F^{(s)}_{\rho}=D_{s,\mathcal{P}}F^{(s)}_{\rho}
\]
for each $s\in\mathbb{R}$ by transforming both sides of the equation with use of Prop.\ref{GenSol3}. Thus we can analyze quantum systems as dynamical systems on coherent phase spaces. As a remark, the problem of finding the differential operator $D_{s,\mathcal{P}}$ is so complexified because it depends on the structure of the operator algebra and the form of $\mathcal{P}$. An actual solution for the 1-degree-of CCR system is seen in \cite{Ban}.

\section{Summary and perspectives}\label{Sum}

In this article we have reviewed general structures of GCS systems mainly based on Perelomov's theory \cite{Per2}, and discussed its application to formulation of QPD's on coherent phase spaces. We have given QPD's operational meanings related to approximately simultaneous measurements, and have constructed concrete formulae of kernel operators. It is remarkable that (K\"{a}hler) geometric structures of coherent phase spaces play essential roles in our discussion; for other words, clearness of our formulation owes to the possibility of direct reciprocity between q- and c-numbers via phase spaces. In addition, the last discussion in \S\ref{GenQPD} propose an answer for the question ``What are coherent states?": Physically, GCS systems are ingredients for a universal semiclassical picture of quantum systems which leads us to the analysis of dynamical equations of QPD's. Their complex structures gives parametrizations and mutual relations of states, and homogeneous spaces as domains of dynamical equations are, in physical words, phase spaces consists of classical variables.

We believe that our scheme has a lot of use for concrete and conceptual investigation of quantum systems in various contexts, for example,
\begin{itemize}
\item A viewpoint as a family of states $\{F^{(s)}(\xi);s\in\mathbb{R}\}$ leads us to entropy-analytic methods,
\item A viewpoint as a model manifold $\{F^{(s)}_{A}(\xi);\hat{A}\in\mathcal{M}\}$ (sets of distribution functions) leads us to information-geometric methods,
\item For the case of $s\neq 1$, the approximately simultaneous measurement scheme shown in Prop.\ref{GenSol3} causes non-positive values, which is deeply related to weak values of observables,
\end{itemize}
and so on. For the former two topics, such a problem is considerable: ``Which $s\in\mathbb{R}$ gives the best model for an actual situation?" It is expected that the factor $\Delta(\xi,\eta)=|K(\xi,\eta)|^{2}$, which is directly derived from the ingredients of GCS systems, plays essential roles in such contexts as well.

\section*{Acknowledgments}

The author owes very important debts to Prof. I. Ojima for his great encouragements and advices, especially on the use of the orbit method, and to Mr. H. Saigo for discussion on symplectic structures in coherent state systems. He is very grateful to Mr. H. Ando, Mr. T. Hasebe, Mr. K. Okamura and Prof. S. Tanimura for their valuable discussions and comments.

\end{document}